\documentclass[letterpaper, 10 pt, conference]{ieeeconf}  

\IEEEoverridecommandlockouts                              

\usepackage{graphicx}
\usepackage{subfigure} 
\usepackage{epsfig} 
\usepackage{mathptmx} 
\usepackage{times} 
\usepackage{amsmath} 
\usepackage{amssymb}  
\usepackage{float}

\title{\LARGE \bf
Comparison of two non-linear model-based control strategies for autonomous vehicles
}

\author{Eugenio Alcal\'{a}, Laura Sellart, Vicen\c{c} Puig, Joseba Quevedo, Jordi Saludes, David V\'{a}zquez and Antonio L\'{o}pez 
\thanks{E. Alcal\'{a}, V. Puig, J. Quevedo and J. Saludes are with the Advanced Control Systems Group, Automatic Control Department, Universitat Polit\`{e}cnica de Catalunya (UPC), Campus de Terrassa,
        Rambla Sant Nebridi, 10, 08222, Spain.
        {\tt\small e-mail: joseba.quevedo@upc.edu}}%
\thanks{L. Sellart, D. V\'{a}zquez and A. L\'{o}pez are with Computer Vision Center (CVC), Edifici O, Campus UAB, 08193 Bellaterra, Barcelona
      }%
}

\begin{document}

\maketitle
\thispagestyle{empty}
\pagestyle{empty}

\begin{abstract}

This paper presents the comparison of two non-linear model-based control strategies for autonomous cars. A control oriented model of vehicle based on a bicycle model is used. The two control strategies use a model reference approach. Using this approach, the error dynamics model is developed. Both controllers receive as input the longitudinal, lateral and orientation errors generating as control outputs the steering angle and the velocity of the vehicle. The first control approach is based on a non-linear control law that is designed by means of the Lyapunov direct approach. The second approach is based on a sliding mode-control that defines a set of sliding surfaces over which the error trajectories will converge. The main advantage of the sliding-control technique is the robustness against non-linearities and parametric uncertainties in the model. However, the main drawback of first order sliding mode is the chattering, so it has been implemented a high order sliding mode control. To test and compare the proposed control strategies, different path following scenarios are used in simulation. \\
\end{abstract}

\section{INTRODUCTION}

Autonomous driving systems have been actively researched. Several works have shown the possibility of autonomous driving in real life \cite{Carvalho2015}. Google has been testing its autonomous vehicle in actual traffic conditions. In August 2012, Google announced that they have completed over 500,000 km autonomous driving without any accident \cite{Urmson2012}. Another research group in Italy, the VisLab in Parma University, did 13,000 km test run for autonomous vehicles from Italy to China \cite{Broggi2010}.
Such a vehicles perceive the surrounding environment by camera sensors and fusion with other sensors. The VisLab also tested the autonomous vehicle in a real environment, together with real traffic on July 2013. In May 2014, Google presented a new concept for their driverless car that had neither a steering wheel nor pedals \cite{Gannes2014}, and unveiled a fully functioning prototype in December 2014 that they planned to test on San Francisco Bay Area roads beginning in 2015. Google plans to make these cars available to the public in 2020 \cite{Broggi2015}.

\begin{figure}[thpb]
    \centering
    \includegraphics[scale=0.41]{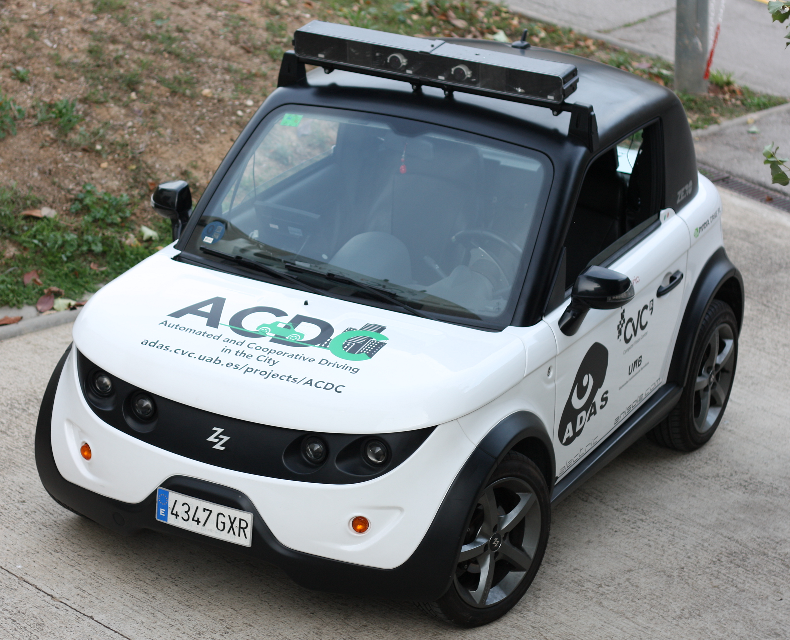}
    \caption{Real autonomous car}
    \label{fig:realcar}
\end{figure}

\begin{figure}[t]
    \centering
    \includegraphics[scale=0.51]{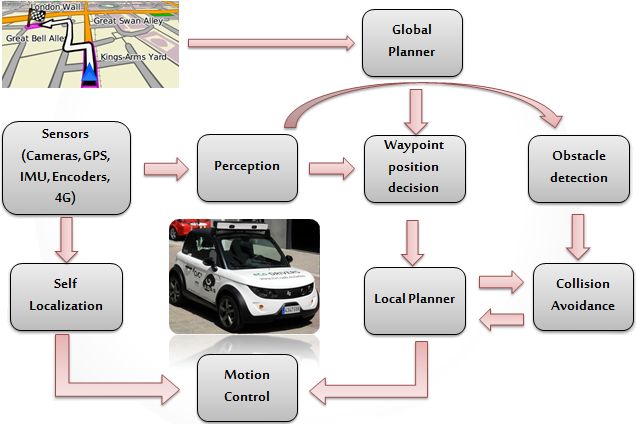}
    \caption{System architecture of the autonomous vehicle}
    \label{architecture}
\end{figure}

The Computer Vision Center (CVC) is automatizing an electric car within the context of the project {\em Automated and Cooperative Driving in the City (ACDC)}\footnote{ http://adas.cvc.uab.es/projects/ACDC/} (see Figure \ref{fig:realcar}).
At this stage, environmental perception is mainly based on Computer Vision. In particular, while following a planned route provided by a global path planner, it is detected the obstacle-free navigable path in front of the vehicle by using an on-board stereo rig.
Accordingly, a short path is planned obtaining the desired set of positions and velocities. Such a set is send to the car controller to properly execute the maneuver.
Figure \ref{architecture} shows the architecture of the project. In this paper, we focus on the design and implementation of such a vehicle low-level controller.

\begin{figure}[thpb]
    \centering
    \includegraphics[scale=0.23]{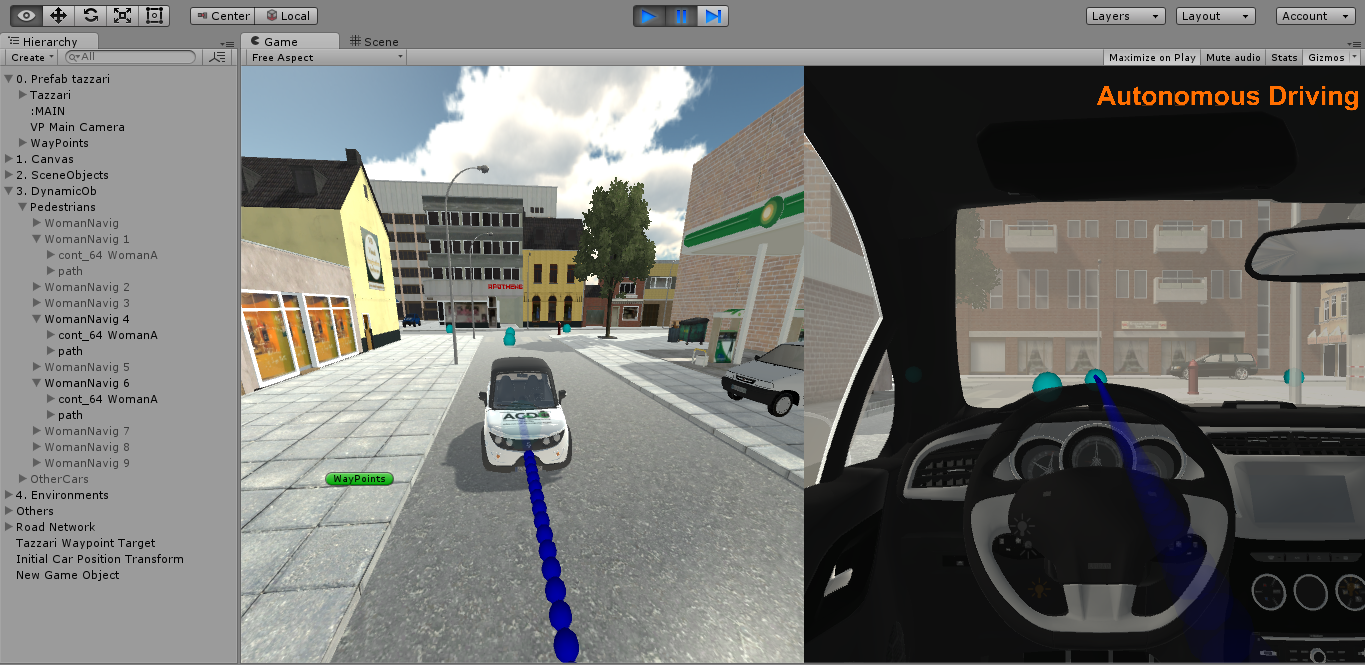}
    \caption{Virtual reality vehicle simulator in Unity}
    \label{fig:unity1}
\end{figure}

\noindent This paper is focused on the low frame of the automatic control of the speed and the steering angle of the car following a predefined path with the best performances of stability and precision. Aware that a good automatic car control is the basis for achieving the other challenges of the autonomous driving systems.
We propose two strategies of non-linear automatic low level control, based on the method of Lyapunov \cite{Dixon2004} and based on sliding mode, and a comparison of both has been made in a simple simulator (based on Simulink). Currently, it is being tested in a complex simulator developed in Unity 3D\footnote{http://unitypackages.net/unitycar/joomla/index.php} (see Figure \ref{fig:unity1}).

This paper is organized as follows. Section II introduces the kinematic model of the vehicle that has been used. In Section III, we develop the control strategies. Section IV presents the results of such strategies. Section V presents the conclusions.


\section{CONTROL ORIENTED VEHICLE MODEL}
\begin{figure}[thpb]
      \centering
      \includegraphics[scale=0.5]{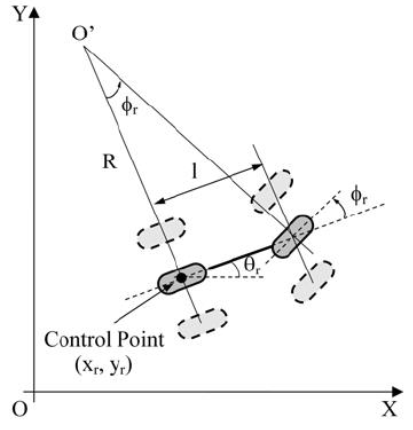}
      \caption{Bicycle model}
      \label{f1}
   \end{figure}

\begin{figure}[thpb]
      \centering
      \includegraphics[scale=0.4]{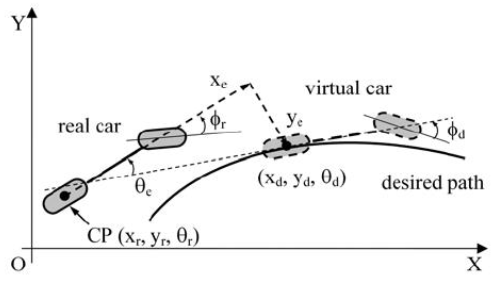}
      \caption{Vehicle's position and orientation with respect to the dynamic target frame (virtual car)}
      \label{f2}
   \end{figure}

For control design, the autonomous car has been considered as a bicycle-like vehicle (\cite{Aicardi1995}, \cite{Solea2007} ) positioned at a non-zero distance with respect to a dynamic waypoint (virtual car of reference), whose motion is controlled by the combined action of both the angular velocity $\omega_{r} (t)$ and the linear velocity $v_{r} (t)$ of the real vehicle (Figures \ref{f1} and \ref{f2}).
This model assumes that the vehicle is symmetric, the steering angle is the same in both front wheels, the roll and pitch movement are neglected, the linear movement in $z$ axis is also neglected and angles like steering and yaw are assumed to be small.
       \\
Then, the set of kinematic equations of the cartesian position ($x_r, y_r$) and orientation ($\theta_{r}$) of the real vehicle is presented as follows:
\begin{equation}
    \begin{cases}
        \dot x_{r} = v_{r} sin(\theta_{r})  \\
        \dot y_{r} = v_{r} cos(\theta_{r}) \\
        \dot \theta_{r} = \frac{ v_{r}}{l} tan(\phi_{r})
    \end{cases}
    \label{eq:real_veh}
\end{equation}
where $v_r$ and $\phi_r$ represent the linear velocity and the steering angle respectively.
Consequently, from Fig. \ref{f2}, the kinematic equations for the virtual car can be defined as:
\\
\begin{equation}
    \begin{cases}
        \dot x_{d} = v_{d} sin(\theta_{d})  \\
        \dot y_{d} = v_{d} cos(\theta_{d}) \\
        \dot \theta_{d} = \frac{ v_{d}}{l} tan(\phi_{d})
    \end{cases}
    \label{eq:ref_veh}
\end{equation}
\\
where $x_d$, $y_d$ and $\theta_d$ are the position and orientation of the next way point generated by the trajectory planner.
\\
The error model (3) is defined as the difference between real vehicle position and the desired one multiplied by the rotation matrix over $z$ axis which is the orthogonal to the road plane:
\\
\begin{equation}
    \left[\begin{array}{c}
     x_{e} \\
     y_{e} \\
    \theta_{e} \\
    \end{array}\right]=
    \left[\begin{array}{ccc}
     cos \theta_{d} & sin \theta_{d} & 0 \\
     -sin \theta_{d} & cos \theta_{d} & 0 \\
     0 & 0 & 1 \\
    \end{array}\right]
    \left[\begin{array}{c}
     x_{r} - x_{d}  \\
     y_{r} - y_{d} \\
     \theta_{d} - \theta_{r} \\
    \end{array}\right]
    \label{eq:errors}
\end{equation}
\\
\noindent that after some algebraic manipulations lead to the following expression:
\\
\begin{equation}
    \begin{cases}
        \dot x_{e} = v_{r} cos(\theta_{e}) + y_{e} \frac{ v_{d}}{l} tan(\phi_{d}) - v_{d}  \\
        \dot y_{e} = v_{r} sin(\theta_{e}) - x_{e} \frac{ v_{d}}{l}  tan(\phi_{d}) \\
        \dot \theta_{e} = \frac{ v_{r}}{l} tan(\phi_{r}) - \frac{ v_{d}}{l}  tan(\phi_{d})
    \end{cases}
    \label{statespace}
\end{equation}
\\

\noindent that can be expressed as follows taking into account the real and reference vehicle models \eqref{eq:real_veh}-\eqref{eq:ref_veh}:
\\
\begin{equation}
    \begin{cases}
        \dot x_{e} = v_{r} + \omega_r y_{e} -v_{d}cos(\theta_{e}) \\
        \dot y_{e} = -\omega_r x_{e}+ v_{d}cos(\theta_{e})\\
        \dot \theta_{e} = \omega_r-\omega_{d}
    \end{cases}
    \label{eq:err_model}
\end{equation}
\\

\noindent where $\omega_r = \dot {\theta_r}$ and $\omega_d = \dot {\theta_d}$.

\section{DESCRIPTION OF AUTOMATIC CONTROL STRATEGIES}
The autonomous car control objective of path following via way points consists to reach asymptotically to zero the difference between the dynamic position and orientation of the real car respect to the dynamic way points (virtual car) position and orientation.

\noindent In this paper, two nonlinear automatic control strategies for autonomous vehicles for path following and navigation among way-points have been considered: one based on the direct method of Lyapunov \cite{Dixon2004} and the other based on sliding mode control (SMC) \cite{Slotine1991}.
Both techniques will consider as starting point the error model derived from the vehicle control-oriented model presented in previous section.

\noindent On one hand, the idea of nonlinear control based on the direct Lyapunov method is to define a control law assuring the stability and the asymptotic elimination of the following error.

\noindent On the other hand, the basic idea of SMC is to reach a sliding surface in finite time and remain on this. However, this control approach has a drawback: the chattering, i.e. a trajectory oscillation over the sliding surface. There are several ways of dealing with this problem e.g., using a higher order sliding surface or smoother functions instead of the common sign function.

\subsection{Direct Lyapunov approach}
We are going to design a control based on the direct Lyapunov approach. This method guaranties the asymptotic stability of the vehicle control because:
\\
\begin{equation}
    \lim_{t\to\infty} \left[\begin{array}{c}
     \dot x_{e} \\
     \dot y_{e} \\
     \dot \theta_{e} \\
    \end{array}\right] = 0
\end{equation}
\\
\noindent which involves also:
\begin{equation}
    \lim_{t\to\infty} \left[\begin{array}{c}
     x_{r} - x_{d}  \\
     y_{r} - y_{d} \\
     \theta_{r} - \theta_{r}  \\
    \end{array}\right] = 0
\end{equation}
\\
\noindent As control law we propose to use the non linear law from \cite{Dixon2004}:
\\
\begin{equation} \label{eq:control_law}
    \left[\begin{array}{c}
     v_{r} \\
     \omega_{r} \\
    \end{array}\right]=
    \left[\begin{array}{c}
     v_{d} cos \theta_{e} - k_{1} x_{e}  \\
     \omega_{d} - k_{2} v_{d} \frac{sin \theta_{e}}{\theta_{e}} y_{e} - k_{3} \theta_{e}  \\
    \end{array}\right]
\end{equation}
\\
Given the following Lyapunov function:
\\
\begin{equation}
V = \frac{1}{2}x_e^2 + \frac{1}{2}y_e^2 + \frac{1}{2}\theta _e^2
\end{equation}

\noindent the stability condition is achieved when $\dot V\leq0$. Thus, taking into account the error model \eqref{eq:err_model}, the proposed control law \eqref{eq:control_law} and substituting in the derivative of (9) and after some simplifications we obtain the following expression:

\begin{equation}
\dot V = -k_1 k_2 x_{e}^2-k_3\theta _e^2 \leq 0
\end{equation}

\noindent which implies that the control parameters $k_1$, $k_2$ and $k_3$  should be positive to assure the asymptotic stability of the closed-loop.

\subsection{Sliding mode control}

The main idea behind this approach is to reach the sliding surface in a finite time and remain on such surfaces where the error is null. From the mathematical point of view a sliding surface is an expression composed by states of the system to be minimized.
\\
   \begin{figure}[thpb]
      \centering
      \includegraphics[scale=0.6]{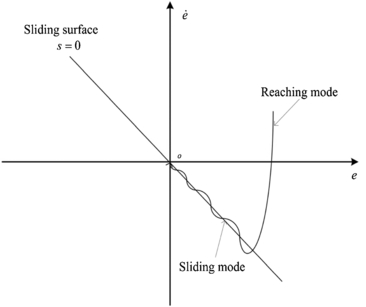}
      \caption{Sliding Mode behaviour}
      \label{sliding}
   \end{figure}
\\
Fig. \ref{sliding} shows the basic principle of the method and it can be seen how the trajectory reaches the surface $s$ and remains there. However, the trajectory presents some oscillations around the surface. This is known as the chattering phenomena. This drawback can be reduced by using a higher order sliding surfaces or using smooth functions instead of the common used sign function.
\\
In this paper, a stable SMC has been designed which has as inputs the errors \eqref{eq:errors}, desired linear velocity ($v_d$), desired angular velocity ($\omega_d$), desired linear acceleration ($\dot v_d$) and desired angular acceleration ($\dot w_d$), and as outputs the velocity ($v_r$) and the steering angle ($\phi_r$).
\\
In order to implement it, a set of sliding surfaces are chosen. There is a surface $s_i$ per control action. Given that there are three error variables ($x_e$, $y_e$ and $\theta_e$) and only two surfaces, one sliding surface have to contain two of such error variables. It has been decided to couple $y_e$ and $\theta_e$ in the same sliding equation, therefore the resulting surfaces are the following:
\\
\begin{equation}
    s_1 = \dot x_{e} + k_1 x_{e}
\end{equation}
\begin{equation}
    s_2 = \dot y_{e} + k_2 y_{e} + k_3 \theta_{e}
\end{equation}
\\
where $k_1$, $k_2$ and $k_3$ are positive defined parameters.
\\
According to \cite{Gao1993}, the dynamics of the sliding surface is the following, which is called the reaching law:
\\
\begin{equation} \label{eq:SMC_law}
    \dot s_i = -Q_i s_i - P_i sign(s_i)
\end{equation}
\\
\noindent where $Q$ and $P$ are positive defined parameters and its stability can be proven using Lyapunov theorem \cite{Slotine1991}.
A Lyapunov candidate function and its time derivative is defined as follows:
\\
\begin{equation}
    V=\frac{1}{2} s' s
\end{equation}

\noindent Evaluating its derivative:
\begin{equation}
    \dot V = s \dot s
\end{equation}

\noindent and considering the control law \eqref{eq:SMC_law}, it can be expressed as:

\begin{equation}
    \dot V = s_1  (-Q_1  s_1 - P_1  sgn(s_1)) + s_2  (-Q_2  s_2 - P_2  sgn(s_2))
\end{equation}

\noindent or alternatively:

\begin{equation}
    \dot V = -Q_1  s_1^2 -Q_2  s_2^2 - P_1 |s_1| - P_2 |s_2|
\end{equation}

\noindent such that to fulfill the Lyapunov stability theorem, $Q_1$, $Q_2$, $P_1$ and $P_2$ have to be semi-positive definite.
\\
Finally, the control law will have the following expression:
\begin{equation} \label{eq:SMC_structure}
    u_i = u_{eq_{i}} - u_{c_{i}}
\end{equation}
\\
where the first term is called equivalent control and makes the derivative of the sliding surface equal to zero to stay on the sliding surface.
\\
The second part of \eqref{eq:SMC_structure} is the corrective control which compensates the deviations from the sliding surface to reach the sliding surface:
\begin{equation}
    \label{eq:SMC_structure_bis}
    u_{c_{i}} = \frac{Q_i  s_i + P_i  sgn(s_i)}{g(x)}
\end{equation}
\\
Hence, in order to obtain the control law it is necessary to find the term $u_{eq_{i}}$. To do so, the sliding surfaces are derived and set equal to zero:
\begin{equation}
    \dot s_1 = \ddot x_{e} + k_1   \dot x_{e}
\end{equation}
\begin{equation}
    \dot s_2 = \ddot y_{e} + k_2  \dot y_{e} + k_3 \dot \theta_{e}
\end{equation}
\\
By developing the two last equations:
\\
\begin{equation}
    \dot s_1 = \dot v_r  cos(\theta_e) + v_r  \dot \theta_e  sin(\theta_e) + \dot y_e  \omega_d + y_e  \dot \omega_d - \dot v_d + k_1  \dot x_e
\end{equation}
\begin{equation}
    \dot s_2 = \dot v_r sin(\theta_e) + v_r cos(\theta_e) \dot \theta_e - x_e \dot \omega_d - \dot x_e \omega_d + K_2 \dot y_e + K_3 \dot \theta_e
\end{equation}
\\
Then, imposing that when reaching the sliding surfaces:

\noindent $\dot s_1 = 0 :$
    \begin{equation}
        u_{eq_{1}} = \dot v_r = \frac{ -v_r \dot \theta_e sin(\theta_e) - \dot y_e \omega_d - y_e \dot \omega_d + \dot v_d - k_1 \dot x_e}{cos(\theta_e)}
    \end{equation}
\\
\noindent $\dot s_2 = 0 :$
    \begin{equation}
        u_{eq_{2}} = \omega_r = \omega_d + \frac{-k_2 \dot y_e + \dot w_{d} x_e + w_{d} \dot x_e  - \dot v_c sin(\theta_e)}{v_r sin(\theta_e) + k_3 }
    \end{equation}
where the denominator corresponds with $g(x)$ in \eqref{eq:SMC_structure_bis}.
\\
Therefore, replacing the obtained equivalent control equations in the structure proposed in \eqref{eq:SMC_structure} and developing them, the following control laws are obtained:
\\
\begin{equation} \label{eq:vel_law_1}
\resizebox{.95\hsize}{!}{$    \dot v_c = \frac{ -v_r \dot \theta_e sin(\theta_e) - \dot y_e \omega_d - y_e \dot \omega_d + \dot v_d - k_1 \dot x_e - Q_1 s_1 - P_1 sgn(s_1)}{cos(\theta_e)}$}
\end{equation}
\\
\begin{equation} \label{eq:vel_law_2}
    v_c = \int_{}^{} \dot v_c  dt
\end{equation}

\begin{equation} \label{eq:steer_law_1}
\resizebox{.95\hsize}{!}{$     \dot \theta_c = w_d + \frac{-P_2 s_2 - Q_2 sign(s_2) - k_2 \dot y_e + \dot w_{d} x_e + w_{d} \dot x_e  - \dot v_c sin(\theta_e)}{v_r sin(\theta_e) + k_3 }$}
\end{equation}
\begin{equation}\label{eq:steer_law_2}
    \phi_c = atan(\frac{l}{v_{r}} \dot \theta_c)
\end{equation}
\\
Equations \eqref{eq:vel_law_1} and \eqref{eq:vel_law_2} represent the velocity control law while \eqref{eq:steer_law_1}  and \eqref{eq:steer_law_2} represent the steering control law.

\section{APPLICATION}

In this section, the results of the previous control methods are presented in simulation.
The simulation has been developed in Matlab/Simulink and it is also currently developed in the Unity platform \footnote{http://unitypackages.net/unitycar/joomla/index.php}.
\\
In parallel with the implementation of the controller, a trajectory planner has also been implemented which provides the specific instructions to the control area.

The next steps are followed to perform the trajectory tracking:
\begin{itemize}
    \item The GPS provides to the vehicle a set of forward way points at every segment. The space between two way points is called segment.
    \item When a segment finishes, the planner takes the next way point and perform the correct speed profile according to the maximum acceleration allowed.
    From this segment a set of sub way points are calculated with its respective position, orientation, linear velocity, angular velocity, linear acceleration and angular acceleration.
    \item Once such a segment has been sampled, at every sample time (Ts = 0.1s) the control area takes a sub way point features as a desired configuration and perform the control.
\end{itemize}

\begin{figure}[htp!]
    \centering
    \includegraphics[scale=0.5]{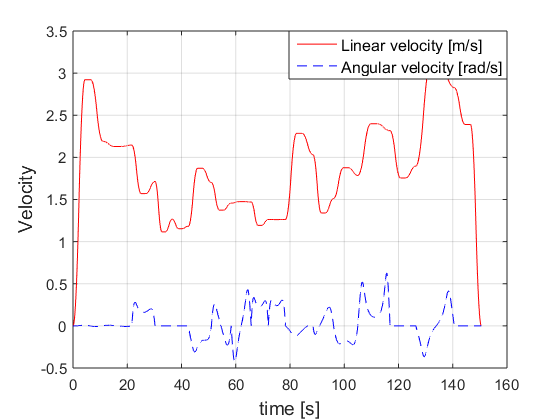}
    \caption{Desired angular and linear velocities}
    \label{speed_profile}
\end{figure}

\begin{figure}[htp!]
    \centering
    \includegraphics[scale=0.5]{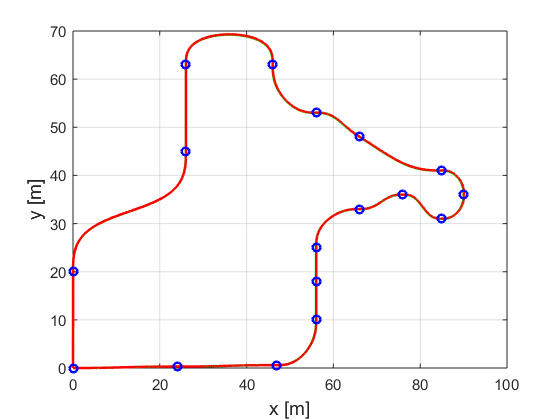}
    \caption{Path proposed to prove control techniques}
    \label{path}
\end{figure}

\noindent Figure \ref{path} shows the total path to be followed, where blue circles represent the way points and between them there are a set of sub way points. Figure \ref{speed_profile} shows the desired set of velocities for the whole path computed by the planner.

\noindent Input disturbances have been included in the sensors data and in the model as gaussian random noises to make the simulation more realistic. Inside the vehicle model, two first order filters have been considered in order to mitigate the high frequency terms of the control signals in case of the sliding model controller.

\noindent In next subsections the results of the control methods are presented.

\subsection{Direct Lyapunov based controller results}
The controller is designed according the procedure described in Section III.A. The design parameters are adjusted as follows: $k_{1}=0.9$, $k_{2}=1.1$  and $k_{3}=3$. They have been adjusted by trial and error. Fig. 10 presents the resulting path, the control actions and the errors. It can be seen that the errors are sufficiently small. For instance in the case of lateral vehicle error the maximum value reached when the car arrives to a curve is 5 cm.

\begin{figure}[thpb!]
    \label{results_Lyapunov}
    \centering
    \subfigure[]{\includegraphics[width=38mm]{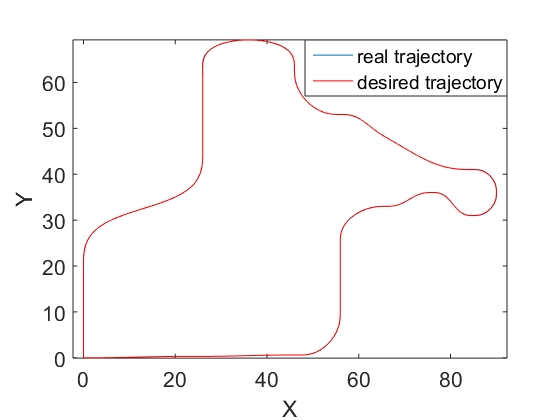}}\hspace{0mm}
    \subfigure[]{\includegraphics[width=38mm]{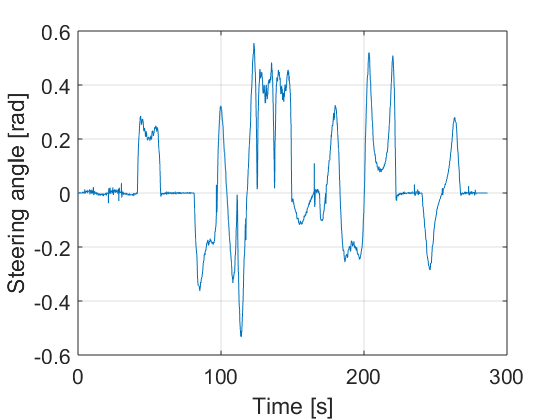}}
    \subfigure[]{\includegraphics[width=38mm]{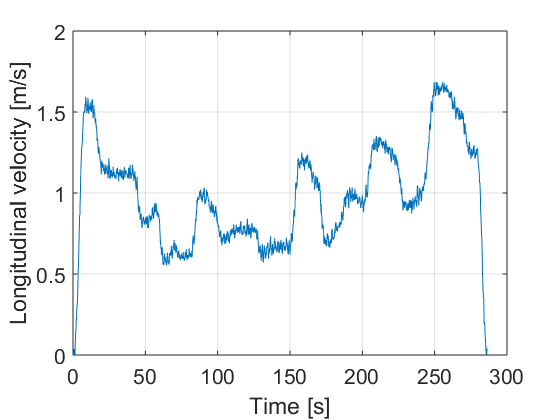}}\hspace{0mm}
    \subfigure[]{\includegraphics[width=38mm]{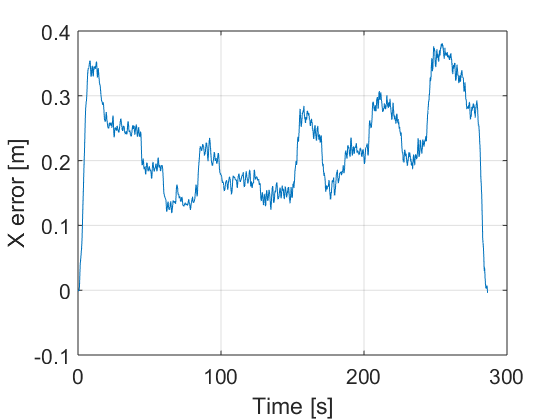}}
    \subfigure[]{\includegraphics[width=38mm]{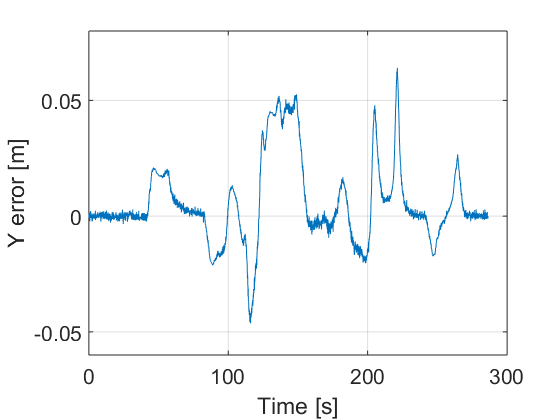}}\hspace{0mm}
    \subfigure[]{\includegraphics[width=38mm]{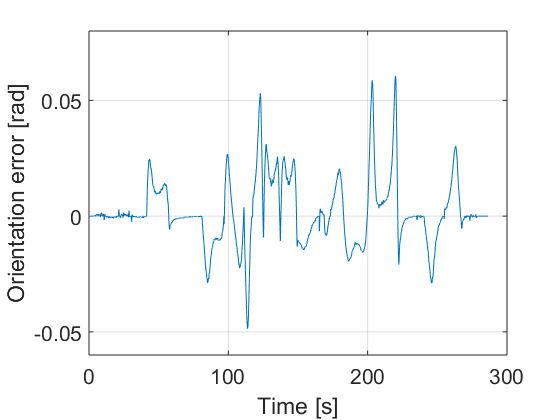}}
    \caption {Direct Lyapunov-based controller results: (a) Desired and tracked path. (b) Steering angle applied to the vehicle. (c) Longitudinal speed. (d) Longitudinal vehicle error. (e) Lateral vehicle error. (f) Orientation vehicle error.}
\end{figure}

\subsection{Sliding mode control results}
The controller is designed according the procedure describes in Section III.B. The set of sliding parameters are shown in Table \ref{SMC_param}, which have been adjusted by trial and error too.

\begin{figure}[thpb]
\centering
\subfigure[]{\includegraphics[width=38mm]{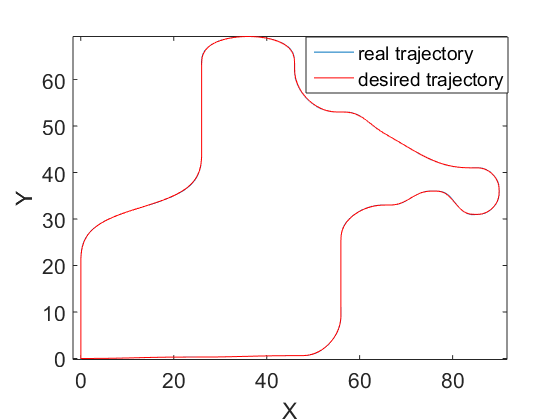}}\hspace{0mm}
\subfigure[]{\includegraphics[width=38mm]{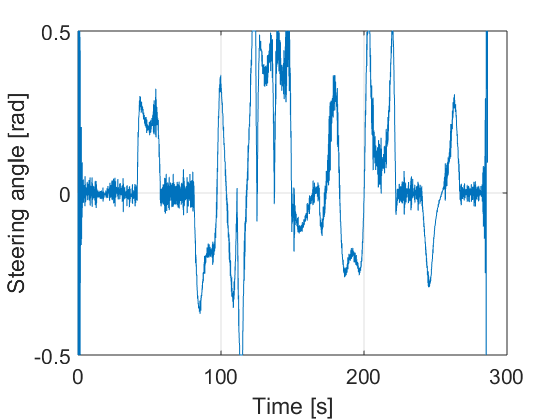}}
\subfigure[]{\includegraphics[width=38mm]{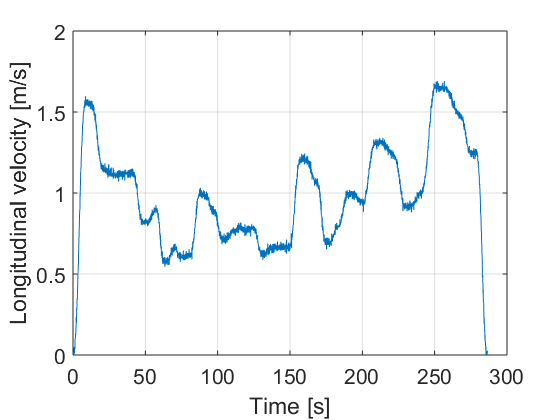}}
\subfigure[]{\includegraphics[width=38mm]{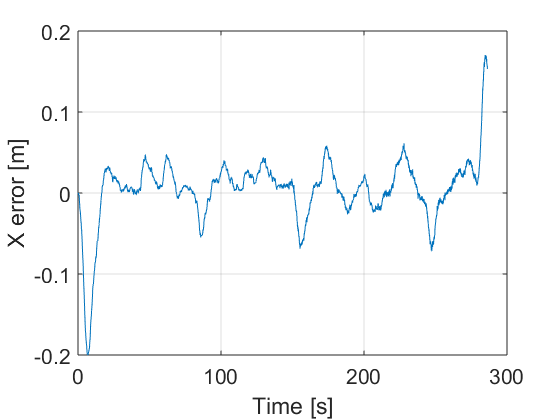}}\hspace{0mm}
\subfigure[]{\includegraphics[width=38mm]{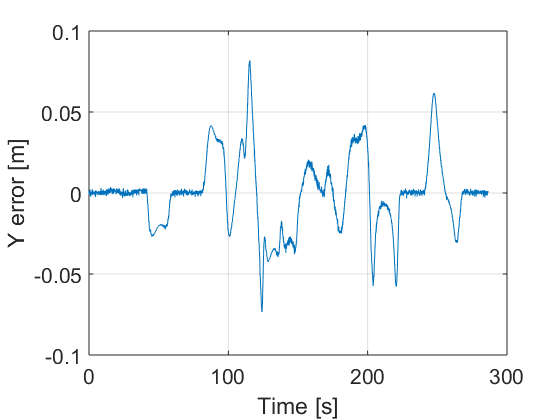}}
\subfigure[]{\includegraphics[width=38mm]{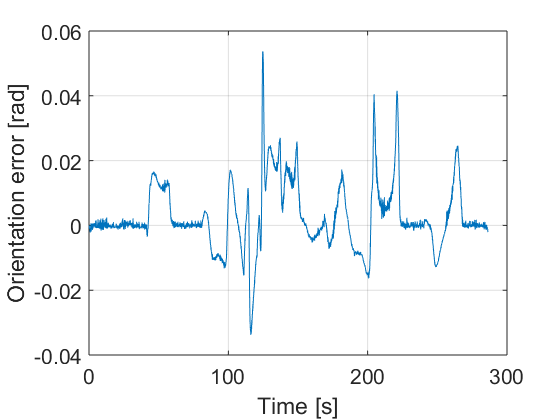}}
\caption{Sliding mode controller results: (a) Desired and tracked path. (b) Steering angle applied to the vehicle. (c) Longitudinal speed. (d) Longitudinal vehicle error. (e) Lateral vehicle error. (f) Orientation vehicle error.}
\label{results_SMC}
\end{figure}

\begin{figure}[thpb]
\centering
\subfigure[]{\includegraphics[width=40mm]{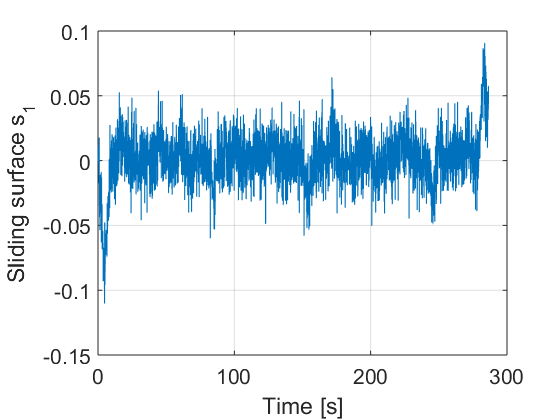}}\hspace{0mm}
\subfigure[]{\includegraphics[width=40mm]{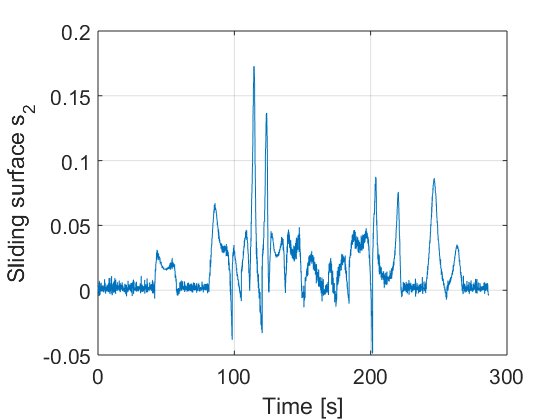}}
\caption{(a) Sliding surface of the longitudinal error. (b) Sliding surface of the lateral and orientation error. }
\label{sliding_surf}
\end{figure}

\noindent Fig. \ref{results_SMC} shows the results using the sliding mode controller.
The vehicle is able to follow the proposed path with small error under noise and perturbation. It can be appreciated in Fig. \ref{results_SMC} d), e) and f) that the errors are quite small.
In fig. \ref{sliding_surf}, the sliding surfaces are showed where it can be seen how they try to reach the zero value and once there remain there.
\\
\begin{table}[thpb!]
\caption{Sliding mode control parameters}
\label{SMC_param}
\begin{center}
\begin{tabular}{|c||c|}
\hline
$k_{1}$ & 0.22\\
\hline
$k_{2}$ & 2\\
\hline
$k_{3}$ & 2.55\\
\hline
$P_{1}$ & 0.48\\
\hline
$Q_{1}$ & 0.048\\
\hline
$P_{2}$ & 3.7\\
\hline
$Q_{2}$ & 0.3\\
\hline
\end{tabular}
\end{center}
\end{table}
\\

\subsection{Result discussion}
By observing the results we can admit both methods provide similar results and they are very good.
\\
Note that the steering angle control action of the Lyapunov technique is smother. It is due to the fact that the SMC method perform the steering signal by using the three errors and the acceleration action, and all these variables have noise. On the other hand, Lyapunov technique only uses two of the three errors to compute the steering angle.
\\
Notice also that the sliding mode method achieves smaller errors although such an errors can be minimised by obtaining a better set of parameters in both techniques.
\\
Both control methods have demonstrated to be robust with respect to some noise and disturbances, and the obtained results show the effectiveness of such proposed control schemes.

\section{CONCLUSIONS}

This work has  presented the comparison of two non-linear model-based control strategies for autonomous cars. Both controllers have been designed using a control oriented model of vehicle based on a bicycle model commonly used in the literature for modelling autonomous cars. The two control strategies follow a model reference approach. Using this approach, the error dynamics model has been developed. Both controllers receive as input the longitudinal, lateral and orientation errors generating as control outputs the steering angle and the velocity of the vehicle. The first non-linear control approach has been designed by means of the Lyapunov direct approach. The second approach has been designed using the sliding mode approach. Both controllers have been implemented, tested and compared with different path following scenarios in simulation. From the obtained results, both methods provide similar results being quite robust to uncertainty and noise.

\noindent As already commented, currently, both control strategies are being tested on a virtual reality simulation developed in Unity before being tested in a real car available at the Computer Vision Center.

\addtolength{\textheight}{-12cm}   

\section*{ACKNOWLEDGEMENTS}

This work is supported by the Spanish MEC project TRA2014-57088-C2-1-R, by Ministerio Economia y Competitivad (MINECO) and FEDER through the project CICYT HARCRICS
DPI2014-58104-R and DGT project SPIP2014-01352, by the Secretaria d'Universitats i Recerca del Departament d'Economia i Coneixement de la Generalitat de Catalunya (2014-SGR-1506). Our research is also kindly supported by NVIDIA Corporation in the form of different GPU hardware.

\end{document}